\documentclass[12pt,a4paper]{article}
\usepackage[utf8]{inputenc}
\usepackage[T1]{fontenc}
\usepackage{amsmath,amssymb,bbm}
\usepackage{graphicx}

\begin{document}
\textwidth=135mm
 \textheight=200mm
\begin{center}
{\bfseries Quark matter in high-mass neutron stars?}
\vskip 5mm
R. Lastowiecki$^\dag$\footnote{Email: lastowiecki@ift.uni.wroc.pl}, D. Blaschke$^{\dag,\ddag}$, 
T. Fischer$^{\dag}$ and T. Kl\"ahn$^{\dag}$
\vskip 5mm
{\small {\it $\dag$ Instytut Fizyki Teoretycznej, Uniwersytet Wroc\l{}awski, 
50-204 Wroc\l{}aw, Poland}}\\
{\small {\it $^\ddag$ Bogoliubov Laboratory for Theoretical Physics, 
JINR, 141980 Dubna, Russia}} \\
\end{center}
\vskip 5mm
\centerline{\bf Abstract}
The recent measurements of the masses of the pulsars PSR J1614-2230 and PSR J0348-0432
provide independent proof for the existence of neutron stars with masses in 
range of 2 $M_\odot$.
This fact has significant implications for the physics of high density matter 
and it challenges the hypothesis that the cores of NS can be composed of deconfined 
quark matter.
In this contribution we study a description of quark matter based on the Nambu--Jona-Lasinio 
effective model and construct the equation of state for matter in beta equilibrium.
This equation of state together with the hadronic Dirac-Brueckner-Hartree-Fock equation of state
is used here to describe neutron star and hybrid star configurations.
We show that compact stars masses of 2 $M_\odot$ are compatible with the possible
existence of deconfined quark matter in their core. 
\vskip 10mm

\section{\label{sec:intro}Introduction}

Neutron stars (NS) are of particular interest for high energy physics due to the fact 
that they probe the high density and low temperature region of the QCD phase diagram which is
otherwise inaccessible for terrestrial experiments.
Studying NS physics can provide sources of information on QCD complementary to those 
from heavy ion collisions experiments.
A critical observable for NS is their maximum mass \cite{Lattimer}, a quantity
that can pose strong restrictions on the high density equation of state (EoS).
A more advanced step forward towards a determination of the full equation of state would involve 
Bayesian analysis of data on mass and radii of NS; significant effort has been devoted to this 
pursuit \cite{Lattimer2, Ayriyan}.

The measurement of two massive NS, PSR J1614-2230 ($M = 1.97\pm0.04 M_{\odot}$)
and PSR J0348-0432 ($M = 2.01\pm0.04 M_{\odot}$) brought new intensity
into the discussion on the implications of high maximum masses of NS
for the EoS of cold and dense matter \cite{Demorest, Antoniadis}.
It was suggested that these measurements put strain on the concept of deconfined
quark matter in NS cores \cite{Lattimer}.
But with the previous work \cite{Klahn} as well as this one we signal that NS with
a mass of $2~M_\odot$ can be compatible with the quark matter hypothesis and we show this
conclusively in the framework of an NJL model description of quark matter.

In this contribution we present results of a systematic scan of the parameter
space of the superconducting three-flavour NJL model, with particular attention given to the compatibility
with high mass measurements for NS.
As a supplementary constraint on  our equation of state we introduce the 
flow constraint from heavy ion collision data \cite{Danielewicz}.
With these two constraints we attempt to narrow down the allowed parameters region for diquark 
and vector couplings in the NJL model at fixed scalar coupling.

\section{\label{sec:NJL}NJL-type models}

The NJL model was created by Nambu and Jona-Lasinio as a way to explain the origin of nucleons mass using an analogy to the superconducting gap in the BCS theory.
In modern use the model is applied to the quark degrees of freedom \cite{Buballa} in order 
to model the chiral symmetry breaking and restoration thus to describing the origin of most of the 
constituent quarks mass.
An advantage of NJL-type models is the possibility for a large number of different interaction 
channels which would result, for example, from a Fierz rearrangement of a global colour model of QCD. \cite{Cahill}
It is nevertheless necessary to note that NJL model misses some of the crucial features of the QCD, 
in particular the local gauge SU(3) symmetry and connected to it the phenomenon of quark confinement \cite{Klahn2}.

From the variety of different channels that can be included in the NJL Lagrangian one needs 
to make a choice of the most relevant ones, in order to keep the model tractable and possible to solve.
The most important, and most standard channel is the scalar channel\footnote{We suppress the explicit notation of the pseudoscalar partner channel required for chiral symmetry of the interaction model Lagrangian since it gives no contribution to the quark mass and the thermodynamics at the meanfield level considered here.}
\begin{equation}
 {\cal L}_S = G_S\sum_{a=0}^{8}(\bar{q}\tau_a q)^2~.
\end{equation}
It is responsible for the chiral symmetry restoration and breaking, and its mean field expectation value
can be closely linked to the quark mass.
The second channel of significant importance is the vector interaction channel
\begin{equation}
 {\cal L}_V = G_V(\bar{q}i\gamma_\mu q)^2~,
\end{equation}
which provides a repulsive interaction between quarks. 
This property is of particular importance for the stiffness
of the resulting quark matter EoS that is essential for describing high-mass hybrid stars.
The third channel that was considered in this work is the diquark interaction channel
\begin{equation}
 {\cal L}_D = G_D\sum_{a,b=2,5,7}
(\bar{q}i\gamma_5\tau_a\lambda_b C \bar{q}^T)(q^TCi\gamma_5\tau_a\lambda_a q)~,
\end{equation}
which is responsible for the formation of colour superconducting condensates and superconducting gaps
in the quark spectrum.
This channel has also a significant impact on the properties of the equation of state as we will
show in the next section.

These three channels taken together with the standard kinematic term
\begin{equation}
{\cal L}_{0} = \bar{q}(-i\gamma^\mu \partial_\mu +\hat{m}+ \gamma_0\hat{\mu})q~,
\end{equation}
building the model Lagrangian that will be studied in the following
\begin{equation}
{\cal L} = {\cal L}_0 + {\cal L}_S  + {\cal L}_V + {\cal L}_D~.
\end{equation}

In order to provide the full description of the EoS, the NJL-model based quark matter
EoS is matched by a Maxwell construction with the well-known hadronic DBHF EoS.
Here we should note that the final results of the calculation of the NS structure
may very well depend on the choice of the hadronic equation of state.

 \section{\label{sec:Results}Results}

\begin{figure}[hbt]
   \centering
   \includegraphics[width=0.95\textwidth,keepaspectratio=true]{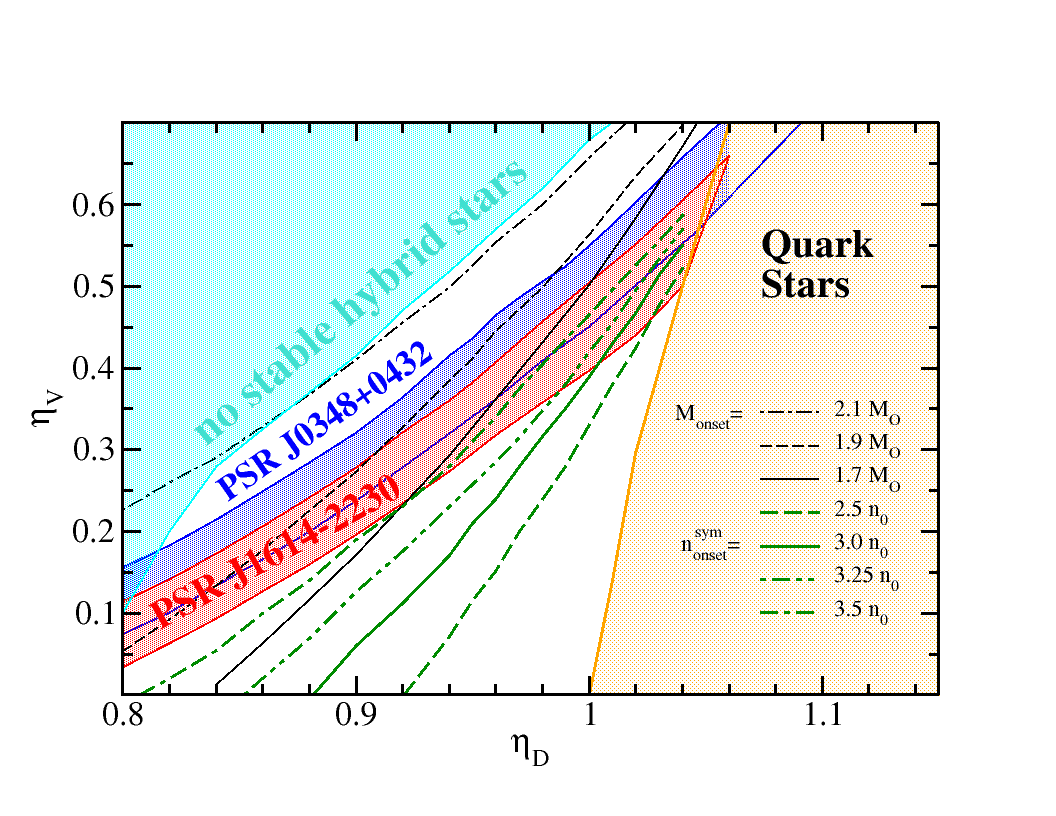}
     \caption{The figure shows the parameter space of the NJL model with marked
		maximum masses and transition densities between DBHF hadronic EoS and NJL quark matter EoS with color superconductivity.}
   \label{Fig:Mon}
 \end{figure}

\begin{figure}[hbt]
   \centering
   \includegraphics[width=0.95\textwidth,height=0.6\textwidth]{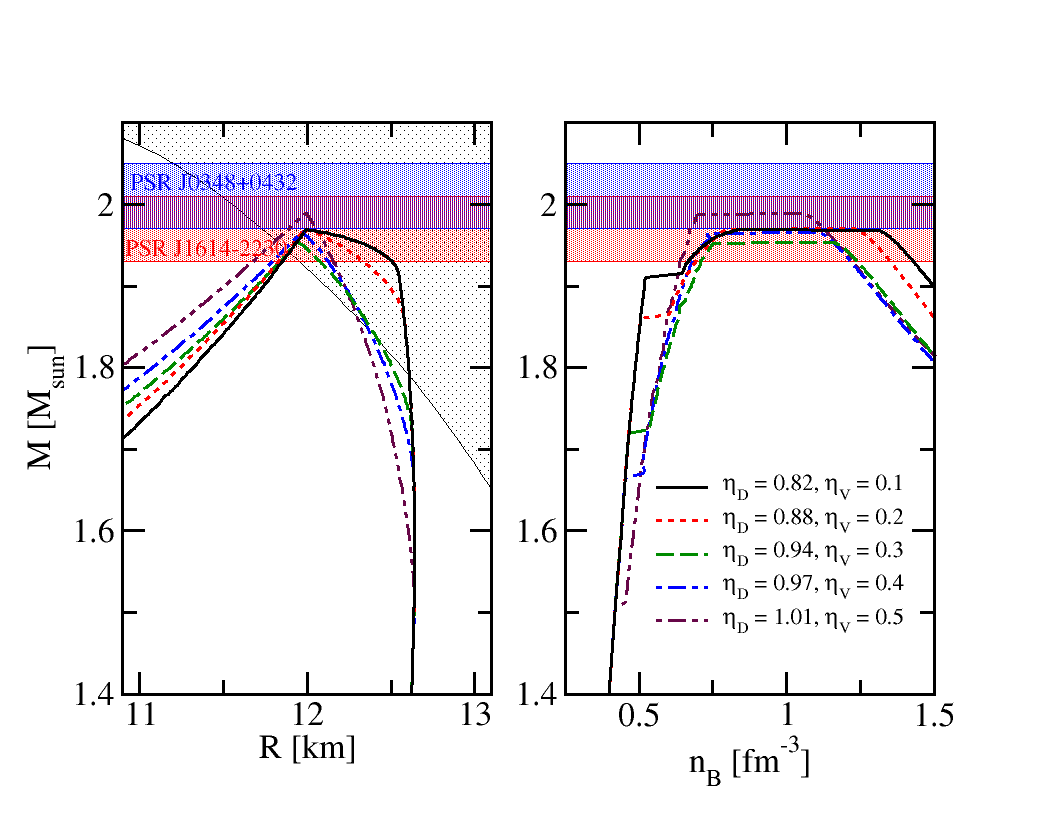}
     \caption{Mass vs. radius (left) and vs. central baryon density (right) for sequences of compact stars
     	corresponding to selected parameter sets of vector and scalar diquark coupling in the quark matter 	Lagrangian, chosen along the constraint for the maximum mass of $1.97~M_\odot$, see 
	Fig.~\ref{Fig:Mon} .}
   \label{Fig:MR}
 \end{figure}

In Fig.~\ref{Fig:Mon} the results of the parameter space scan are presented, the two axes correspond to
the vector and the diquark coupling constants, respectively.
We can note from the graph, that the region of parameters that allow for sufficiently massive hybrid stars
is a rather narrow strip going diagonally through the parameter space.
For small values of the vector coupling also the diquark coupling needs to be sufficiently small in order to produce massive hybrid stars and conversely for large vector coupling the diquark coupling needs to be large enough.
For that reason the most promising parameter region for successful NS phenomenology is for the diquark couplings around $1.0$ and the vector couplings above $0.6$.

In Fig.~\ref{Fig:MR} we show the results for the mass-radius curves of the NS configurations that produce maximum mass high enough to explain the $2 M_\odot$  NS.
We can observe that some of the solutions, although they produce sufficiently massive stars, do not predict presence of  deconfined quark matter in the cores of NS.
Some other solutions allow only for marginal, potentially undetectable amounts of quark matter in the NS. 
Of particular interest for us are the parameter sets that predict both high maximum mass and 
early onset of quark matter, which leads to significant fraction of the stars being composed
of deconfined quarks.
An example of such a promising solution is set with $\eta_D = 1.01$ and $\eta_V = 0.5$,
with an onset of quark matter already in stars with $M\sim 1.5 M_\odot$.
 
\section{\label{sec:Con}Conclusions}
The measurement of high masses of NS around $2~M_\odot$ do not exclude the possibility of the existence of deconfined quark matter in their interiors.
It is nonetheless the case, that these measurements put strong constraints on the possible values of the otherwise free parameters of the NJL model: the diquark and vector coupling constants.
Together with the flow constraint from heavy ion collisions, a narrow region in the parameter space can be identified, where the hypothesis of deconfined quark matter in the NS cores is fulfilled, the maximum mass of two solar masses is reached and the EoS does not violate the flow constraint.
Further studies on the extensions of the NJL model are advised to identify other interaction channels
that could quantitatively change the picture obtained with the present NJL model EoS.

\vskip 10mm
\centerline{\bf Acknowledgment}
R.L. acknowledges support by the Helmholtz Association via the "HISS Dubna" 
programme by the Bogoliubov-Infeld programme for exchange between JINR Dubna and Polish institutes and by the Polish MNiSW under grant no. 1009/S/IFT/14.
This work was supported in part by the Polish National Science Center (NCN) under grant number UMO-2011/02/A/ST2/00306 (D.B. and T.F.) and under grant number UMO-2013/09/B/ST2/01560 (T.K.). 
The authors are grateful for support from the COST Action MP1304 "NewCompStar" for their networking and collaboration activities. 

\end{document}